\documentclass[aps,preprint,showpacs]{revtex4}%
\usepackage{amsfonts}
\usepackage{amsmath}
\usepackage{amssymb}
\usepackage{graphicx}%
\begin{document}
\preprint{HEP/123-qed}
\title{Influence of the Characteristics of the STM-tip on the Electroluminescence Spectra}
\author{M.~D.~Croitoru$^{a}$, V.~N.~Gladilin$^{a}$, V.~M.~Fomin$^{a,b}$,
J.~T.~Devreese$^{b}$\thanks{E-mail: jozef.devreese@ua.ac.be; tel.: +323 820
2460; fax: +323 820 2245}}
\affiliation{Theoretische Fysica van de Vaste Stoffen (TFVS), Universiteit Antwerpen, Belgium}
\author{M. Kemerink, P. M. Koenraad, J. H. Wolter}
\affiliation{COBRA Inter-University Research Institute, Eindhoven University of Technology,
The Netherlands}

\begin{abstract}
We analyze the influence of the characteristics of the STM-tip (applied
voltage, tip radius) on the electroluminescence spectra from an
STM-tip-induced quantum dot taking into account the many-body effects. We find
that positions of electroluminescence peaks, attributed to the electron-hole
recombination in the quantum dot, are very sensitive to the shape and size of
the confinement potential as determined by the tip radius and the applied
voltage. A critical value of the tip radius is found, at which the
luminescence peak positions as a function of the tip radius manifest a
transition from decreasing behavior for smaller radii to increasing behavior
for larger radii. We find that this critical value of the tip radius is
related to the confinement in the lateral and normal direction.

\end{abstract}
\date{\today  }
\startpage{1}
\endpage{2}
\maketitle

\section{Introduction}

Scanning-tunnelling microscopy techniques have developed into an important
tool for studying semiconductor nanostructures in recent years. The
scanning-tunnelling microscope (STM) covers a wide field of applications. An
example of such an application is the so-called tip-induced quantum dot, a
special and unique type of quantum dot that can only be studied using a STM
\cite{wenderot,Dombrowski,Yamauchi,Kemerink}. When a negative bias is applied
between the metallic STM-tip and the semiconductor sample, the electric field
extends into the structure, and a hole accumulation layer is formed at the
surface under the tip. Such an accumulation layer effectively screens the
electric field. For sufficiently small tip apex radii, quantization occurs
both in the radial direction and along the axis normal to the surface of the
structure. It should be noted that the tip-induced band bending confines only
a single type of carriers, holes in our case. The electrons are repelled from
the region under the tip. By putting a barrier just below the surface,
injected electrons are confined to the surface layer and radiative
recombination with holes in the dot becomes possible.

In this paper we analyze the influence of the characteristics of the
confinement potential, determined by the STM-tip radius and the applied
voltage, on the electroluminescence spectra of a tip-induced quantum dot in
the GaAs layer of the GaAs/Al$_{0.25}$Ga$_{0.75}$As/In$_{0.25}$Ga$_{0.75}$As
quantum well structure \cite{Kemerink}.

\section{Model}

In our model we consider a multilayer structure sketched in Fig. 1. The shape
of the STM tip is modelled by a\ truncated cone because,\ when the tip made of
a soft metal (like Pt in our case)\ is pressed to the semiconductor layer, it
becomes flat. The STM-tip and the GaAs layer (cap layer) are separated from
each other by an insulator layer of width $l_{1}$. The holes and electrons are
confined in the cap layer of width $l_{2}$. The width of the Al$_{0.25}%
$Ga$_{0.75}$As barrier is $l_{3}$. The Al$_{0.25}$Ga$_{0.75}$As layer borders
on the In$_{0.25}$Ga$_{0.75}$As quantum well. The contact area of the STM-tip
with the structure is a circle of radius $R_{\mathrm{tip}}$. Owing to the
axial symmetry of the system, cylindrical coordinates $r$, $\varphi,$ $z$ are
used. The $z$-axis is directed along the symmetry axis of the tip, which is
chosen parallel to the normal to the surface of the structure. While the hole
motion is confined in all directions by the potential well, the electron
motion is confined only along the $z$-axis.

In order to find energy spectra of holes, trapped in the quantum dot, we solve
self-consistently the Poisson equation
\begin{equation}
\frac{1}{r}\frac{\partial}{\partial r}\left(  r\frac{\partial V\left(
r,z\right)  }{\partial r}\right)  +\frac{\partial^{2}V\left(  r,z\right)
}{\partial z^{2}}=-\frac{\rho_{h}\left(  r,z\right)  }{\varepsilon
_{0}\varepsilon_{i}}, \label{Poisson}%
\end{equation}
where $\rho_{h}\left(  r,z\right)  \equiv en_{h}\left(  r,z\right)  $ is the
hole charge density and $\varepsilon_{i}$ is\ the dielectric constant of the
material, together with the Schr\"{o}dinger equation governing the hole
motion:%
\begin{multline}
-\frac{\hbar^{2}}{2}\left(  \frac{1}{m_{j}^{||}\left(  z\right)  }\frac{1}%
{r}\frac{\partial}{\partial r}\left(  r\frac{\partial}{\partial r}\right)
+\frac{1}{m_{j}^{||}\left(  z\right)  }\frac{1}{r^{2}}\frac{\partial^{2}%
}{\partial\varphi^{2}}+\frac{\partial}{\partial z}\frac{1}{m_{j}^{\bot}\left(
z\right)  }\frac{\partial}{\partial z}\right)  \Psi_{j,s,n,m}(r,\phi,z)\\
+V(z,r)\Psi_{j,s,n,m}(r,\phi,z)=E_{j,s,n,m}\Psi_{j,s,n,m}(r,\phi,z),\nonumber
\end{multline}

\begin{equation}
m_{j}^{||,\bot}\left(  z\right)  =\left\{
\begin{array}
[c]{l}%
m_{j}^{||,\bot\left[  GaAs\right]  },\;\mathrm{if}\;z\in\mathrm{GaAs}\\
m_{j}^{||,\bot\left[  Al_{0.25}Ga_{0.75}As\right]  },\;\mathrm{if}%
\;z\in\mathrm{Al}_{0.25}\mathrm{Ga}_{0.75}\mathrm{As.}%
\end{array}
\right.
\end{equation}
Here the index $j$ labels the hole band type: $j=1$ for a hole, which is heavy
for the motion along the $z$-axis and light in the plane of the GaAs layer and
$j=2$ for a hole, which is light for the motion along the $z$-axis and heavy
in the plane of the GaAs layer. For the case of a quantum well, the light and
heavy holes exactly decouple at $\mathbf{k}_{\parallel}=0$. Since the lateral
size of the quantum dot is much larger than the GaAs unit cell, only states
around $\mathbf{k}_{\parallel}=0$ contribute to the wavefunctions in the
quantum dot. The effect of coupling between the valence bands is very weak in
the close vicinity of the center of the Brillouin zone $\mathbf{k}_{\parallel
}=0$ (see, e.g., Refs. \cite{BN,koen}). The index $s$ labels subbands due to
the size quantization of the hole degree of freedom along the $z$-axis. The
index $n$ is the radial quantum number and $m$ is the angular quantum number
of the lateral (in the plane of the GaAs layer) motion of the hole. The
kinetic energy of the hole motion along the $z$-axis is larger than the
kinetic energy of the lateral motion. (Our calculations show that the charge
of holes is concentrated in a relatively thin ($\sim3~\mathrm{nm}$) layer near
the interface insulator/GaAs. The radial size of the smallest dot considered
in this paper is $12~\mathrm{nm}$.) Hence the adiabatic approach can be used,
assuming that the hole degree of freedom along the $z$-axis is "fast" while
the lateral one is "slow". Therefore, the hole wave function is represented in
a product form: $\Psi_{j,s,n,m}(r,\phi,z)=\Psi_{j,s}^{\bot}(z;r)\Psi
_{j,s,n,m}^{||}(r)e^{im\phi},$ where $\Psi_{j,s}^{\bot}(z;r)$ and
$\Psi_{j,s,n,m}^{||}(r)e^{im\phi}$ are the wave functions describing the
motion along the $z$-axis and the lateral motion, respectively. The
Schr\"{o}dinger equation governing $\Psi_{j,s}^{\bot}(z;r)$\ reads%
\begin{equation}
-\frac{\hbar^{2}}{2}\frac{\partial}{\partial z}\frac{1}{m_{j}^{\bot}\left(
z\right)  }\frac{\partial}{\partial z}\Psi_{j,s}^{\bot}(z;r)+V(r,z)\Psi
_{j,s}^{\bot}(z;r)=E_{j,s}\left(  r\right)  \Psi_{j,s}^{\bot}(z;r),
\label{Sch}%
\end{equation}
where $V(z,r)=eU(z,r)+V_{\mathrm{bar}}^{h}(z)$ with $U(z,r)$, the
electrostatic potential. $V_{\mathrm{bar}}^{h}(z)$ represents band offsets for
a hole \cite{Har}.

We solve the Schr\"{o}dinger equation (\ref{Sch}) using the method described
in Ref. \cite{Bal,my}. It yields the spectrum of the subband energies
$E_{j,s}\left(  r\right)  $ and wavefunctions $\Psi_{j,s}^{\bot}(z;r)$. Each
of the energies $E_{j,s}\left(  r\right)  $\ determines the top of a hole
subband in the structure and plays the role of an adiabatic potential for the
lateral motion.

Consequently, the equation, which describes the lateral motion, is as follows:%
\begin{equation}
-\frac{\hbar^{2}}{2m_{j}^{||}\left(  z\right)  }\left(  \frac{1}{r}%
\frac{\partial}{\partial r}\left(  r\frac{\partial}{\partial r}\right)
-\frac{m^{2}}{r^{2}}\right)  \Psi_{j,s,n,m}^{||}(r)+E_{j,s}\left(  r\right)
\Psi_{j,s,n,m}^{||}(r)=E_{j,s,n,m}\Psi_{j,s,n,m}^{||}(r). \label{Sch_L}%
\end{equation}
After solving this equation as described in Ref. \cite{my}, we obtain the
energy spectrum of the holes, confined in the STM-tip-induced quantum dot.

Given a particular number of holes in the quantum dot, we find the quasi-Fermi
level $E_{F}^{h}$\ for holes from the equation:
\begin{equation}
N_{h}=\sum_{j=1,2}~\sum_{s,n,m}\frac{2}{\exp\left(  \frac{E_{j,s,n,m}%
-E_{F}^{h}}{kT}\right)  +1}.
\end{equation}
Then, within the Hartree approximation scheme, the hole density may be
extracted from
\begin{equation}
n_{h}\left(  r,z\right)  =\sum_{j=1,2}~\sum_{s,n,m}\left\vert \Psi
_{j,s,n,m}(r,\phi,z)\right\vert ^{2}\frac{2}{\exp\left(  \frac{E_{j,s,n,m}%
-E_{F}^{h}}{kT}\right)  +1}.
\end{equation}
This is the basic formula employed to invoke a self-consistent solution to the
Schr\"{o}dinger and Poisson equations.

In order to include in our scheme the exchange effects we use the local
density approximation (LDA) as described in \cite{DFT,Jones,Kohn}
\begin{equation}
E_{\mathrm{xc}}^{\mathrm{LDA}}\left[  n\right]  =\left.
%TCIMACRO{\dint }%
%BeginExpansion
{\displaystyle\int}
%EndExpansion
\mathrm{d}\mathbf{r}~e_{\mathrm{xc}}^{\hom}\left(  n_{0}\right)  \right\vert
_{n_{0}\longrightarrow n\left(  \mathbf{r}\right)  },
\end{equation}
where $e_{\mathrm{xc}}^{\hom}$ is the exchange energy of the homogeneous gas
per particle.

The exchange energy of a hole gas was evaluated in Ref. \cite{Bobbert}. The
result for exchange energy per hole is
\begin{equation}
e_{\mathrm{xc}}^{\hom}\left(  n\right)  =\frac{E_{\mathrm{xc}}^{\mathrm{LDA}}%
}{N}=-~\frac{e^{2}}{4\pi\varepsilon_{0}\varepsilon}~\frac{3}{4\pi}\left(
3\pi^{2}n\right)  ^{\frac{1}{3}}\zeta\left(  w\right)  ,
\end{equation}
where the numerical function $\zeta\left(  w\right)  $ is given by%
\begin{equation}
\zeta\left(  w\right)  =\frac{w^{4}+3w^{3}+3w+1-\frac{3}{4}\left(
1-w^{2}\right)  ^{2}\ln\left\vert \frac{1+w}{1-w}\right\vert +\frac{3}%
{4}\left(  1-w^{4}\right)
%TCIMACRO{\dint \limits_{w}^{1}}%
%BeginExpansion
{\displaystyle\int\limits_{w}^{1}}
%EndExpansion
\frac{\mathrm{d}x}{x}\ln\left\vert \frac{1+x}{1-x}\right\vert }{4\left(
1+w^{3}\right)  ^{4/3}}.
\end{equation}
A parameter $w$ represents the ratio between the light- and heavy-hole Fermi
wave vectors:%
\begin{equation}
w\equiv\frac{k_{\mathrm{Fl}}}{k_{\mathrm{Fh}}}=\sqrt{\frac{m_{l}}{m_{h}}.}%
\end{equation}
In the local density approximation, the exchange potential is related to the
exchange energy $e_{\mathrm{xc}}^{\hom}\left(  n\right)  $ by \cite{Kohn}%
\begin{equation}
V_{\mathrm{xc}}\left(  \mathbf{r}\right)  =\frac{1}{e}\frac{\mathrm{d}%
}{\mathrm{d}n}\left[  n~e_{\mathrm{xc}}^{\hom}\left(  n\right)  \right]
\left(  \mathbf{r}\right)  .
\end{equation}
The exchange potential is thus%
\begin{equation}
V_{\mathrm{xc}}\left(  r,z\right)  =-\frac{e}{4\pi\varepsilon_{0}%
\varepsilon_{\mathrm{GaAs}}}~\left(  \frac{3}{\pi}\sum_{j=1,2}\sum
_{s,n,m}\left\vert \Psi_{j,s,n,m}(r,\phi,z)\right\vert ^{2}\right)  ^{\frac
{1}{3}}\zeta\left(  w\right)  , \label{final}%
\end{equation}
This formula is employed to include exchange corrections within LDA in the
self-consistent solution of the Schr\"{o}dinger and Poisson equations. When
$V(r,z)$ is substituted by $V(r,z)+V_{\mathrm{xc}}\left(  r,z\right)  $, we
need a straightforward numerical solver as shown in Fig. 2. The iteration
procedure, involved to calculate the hole states and the corresponding
electrostatic potential is subdivided into four steps. In the first step, the
electrostatic potential $V(r,z)$ in the whole structure is calculated in the
absence of holes. Then the obtained potential is used as the initial guess in
the iteration procedure. In the second step, the Schr\"{o}dinger equation,
describing the hole motion is solved. The obtained wave functions are used to
calculate the hole density. In the third step, solving the Poisson equation
with the charge density obtained in the previous step yields a new
approximation for the electrostatic potential, which is involved in the next
iteration. In the fourth step, we use the charge density obtained as a result
of the second step to calculate the exchange potential Eq. (\ref{final}),
which is also used in the next iteration. At a given number of holes in the
tip-induced quantum dot, the second, the third and the fourth steps are
repeated consecutively. This sequence of steps continues till the maximal
absolute value of the difference between the values of the electrostatic
potential at consecutive iterations becomes less than a certain threshold
($\lesssim0.1~\mathrm{mV}$), which establishes the measure of the accuracy.
When increasing the number of holes in the quantum dot, the electrostatic
potential, obtained earlier for a smaller number of holes, is used as an
initial guess for calculations. This procedure guarantees a continuous link
between the solutions obtained at the consecutive numbers of holes.

Given the electrostatic potential formed by the hole charge, we solve the
Schr\"{o}dinger equation for electrons,%
\begin{multline}
-\frac{\hbar^{2}}{2}\left(  \frac{1}{m_{e}^{||}\left(  z\right)  }\frac{1}%
{r}\frac{\partial}{\partial r}\left(  r\frac{\partial}{\partial r}\right)
+\frac{1}{m_{e}^{||}\left(  z\right)  }\frac{1}{r^{2}}\frac{\partial^{2}%
}{\partial\varphi^{2}}+\frac{\partial}{\partial z}\frac{1}{m_{e}^{\bot}\left(
z\right)  }\frac{\partial}{\partial z}\right)  \Psi_{s_{e},n_{e},m_{e}}%
^{e}(r,\phi,z)\\
+V^{e}(z,r)\Psi_{s_{e},n_{e},m_{e}}^{e}(r,\phi,z)=E_{s_{e},n_{e},m_{e}}%
^{e}\Psi_{s_{e},n_{e},m_{e}}^{e}(r,\phi,z),
\end{multline}
where $V^{e}(z,r)=-eU(z,r)+V_{\mathrm{bar}}^{e}(z)$, $U(z,r)$ is the
electrostatic potential, and $V_{\mathrm{bar}}^{e}(z)$ describes band offsets
for an electron \cite{Har},%

\begin{equation}
m_{e}^{||,\bot}\left(  z\right)  =\left\{
\begin{array}
[c]{l}%
m_{e}^{||,\bot\left[  GaAs\right]  },\;\mathrm{if}\;z\in\mathrm{GaAs}\\
m_{e}^{||,\bot\left[  Al_{0.25}Ga_{0.75}As\right]  },\;\mathrm{if}%
\;z\in\mathrm{Al}_{0.25}\mathrm{Ga}_{0.75}\mathrm{As.}%
\end{array}
\right.
\end{equation}
The index $s_{e}$ labels subbands due to the size quantization of the electron
motion in the direction normal to the surface of the structure. The index
$n_{e}$ is the radial quantum number and $m_{e}$ is the angular quantum number
of the lateral motion of the electron. The electron wave function is
represented in a product form: $\Psi_{s_{e},n_{e},m_{e}}^{e}(r,\phi
,z)=\Psi_{s_{e}}^{\bot,e}(z;r)\Psi_{s_{e},n_{e},m_{e}}^{||,e}(r)e^{im_{e}\phi
}$, where $\Psi_{s_{e}}^{\bot,e}(z;r)$ and $\Psi_{s_{e},n_{e},m_{e}}%
^{||,e}(r)e^{im_{e}\phi}$ are the wave functions describing the motion along
the $z$-axis and the lateral motion, respectively.

Our numerical approach to solve the Schr\"{o}dinger equation, which governs
the lateral degree of freedom of the electron, is described in Ref. \cite{my}.
The set of energies $E_{s_{e},n_{e},m_{e}}^{e}$ and wave functions
$\Psi_{s_{e},n_{e},m_{e}}^{e}(r,\phi,z)$ together with the set of hole
energies $E_{j,s,n,m}$ and wave functions $\Psi_{j,s,n,m}(r,\phi,z)$ are used
to describe the electron-hole radiative recombination. The intensity of
electroluminescence at a frequency $\Omega$\ in the quantum dot created by the
STM tip is described by the following expression \cite{BN}:%
\begin{equation}
I\left(  \hbar\Omega\right)  \sim\sum_{j,s,n,m}~\sum_{s_{e},n_{e},m_{e}}%
f_{h}\left(  j,s,n,m\right)  f_{e}\left(  s_{e},n_{e},m_{e}\right)  P\left(
j,s,n,m,s_{e},n_{e},m_{e},\hbar\Omega\right)  , \label{inten}%
\end{equation}
where $f_{h}\left(  j,s,n,m\right)  $ and $f_{e}\left(  s_{e},n_{e}%
,m_{e}\right)  $ are, respectively, probabilities of finding holes and
electrons in the corresponding quantum states, $P\left(  j,s,n,m,s_{e}%
,n_{e},m_{e},\hbar\Omega\right)  $ is the probability of the electron-hole recombination.

We assume that the distribution functions $f_{h}\left(  j,s,n,m\right)  $ is
described by the Fermi-Dirac function with the quasi-Fermi level $E_{F}^{h}$ :%
\begin{equation}
f_{h}\left(  j,s,n,m\right)  =\frac{1}{\exp\left(  \frac{E_{j,s,n,m}-E_{F}%
^{h}}{kT}\right)  +1}.
\end{equation}
Following Ref. \cite{my}, we assume that due to the radial accelerating field
the non-equilibrium electron distribution is a strongly decreasing function of
the angular quantum number$\ m_{e}$. Hence, the electron distribution function
is approximated as%

\begin{equation}
f_{e}\left(  s_{e},n_{e},m_{e}\right)  \sim\delta_{m_{e}0}. \label{distr2}%
\end{equation}

\section{Results and discussion}

The characteristics of the structure under consideration are as follows. The
thickness of the vacuum barrier is $l_{1}=0.5~\mathrm{nm}$, while the
thickness of the GaAs and Al$_{0.25}$Ga$_{0.75}$As layers are,
correspondingly, $l_{2}=17~\mathrm{nm}$ and $l_{3}=46~\mathrm{nm}$. The
respective dielectric constants are $\varepsilon_{3}=1.0$, $\varepsilon
_{2}=13.2$ and $\varepsilon_{3}=12.2$.

Figure 3 illustrates the position of the calculated peaks in the
electroluminescence spectrum as a function of the voltage $V_{\mathrm{tip}}$
applied between the STM-tip and the semiconductor structure at $4.2~\mathrm{K}%
$. Clearly, the peak positions as a function of the bias between the sample
and the tip reveal a red shift. This is explained as follows. An increase of
the voltage shifts the bottom (top) of the dot (antidot) upwards nearly
proportional to the applied voltage. Since the number of holes increases with
the voltage, the width of the dot in the $z$-direction is reduced due to a
stronger screening. Hence, an increase of $V_{\mathrm{tip}}$ leads to a
deepening of the potential well for holes and to a rise of the barrier for
electrons. This, in its turn, raises the energy of the electron states, which
contribute to the recombination, with respect to the bottom of the conduction
band in the absence of the applied voltage. The holes experience both the rise
of the energy level due to the shift of the bottom of the dot and the lowering
of the energy level due to a stronger confinement. The net effect of both
trends is an upward shift of the hole energy levels. The upward shift of the
hole energy levels with increasing $V_{\mathrm{tip}}$ is larger than that for
electrons. Thus, these shifts result in a \textit{red shift} in the
electroluminescence spectrum (see also Ref. \cite{my}).

In Fig. 4 we present the peak positions for the spectrum shown in the inset of
Fig. 3 as a function of the radius of the STM-tip. When the radius of the
STM-tip $R_{\mathrm{tip}}>12\ \mathrm{nm}$, one notices that the enlargement
of the radius $R_{\mathrm{tip}}$\ leads to a rise of the transition energies
(blue shift). This is understood on the basis of the fact that an enlargement
of the contact area shifts on average the conduction and valence bands
upwards, resulting in a rising potential barrier for electrons and a slight
deepening of the potential well for holes. Consequently, the recombining
electron has to occupy a higher energy level in order to have a considerable
overlap integral with a hole wave function. Our calculations show that the
upward shift of the hole levels is smaller than the increase of the energy of
the recombining electron. So, a change of the contact area influences both
hole and electron levels, but due to the weak confinement for holes in the
radial direction, the shifts of the electron and the hole levels jointly lead
to a \textit{blue shift} of the transition energy (this behavior of the
transition energy as a function of the tip radius has been analyzed in Ref.
\cite{my}).

As distinct from that type of behavior, for $R_{\mathrm{tip}}<12\ \mathrm{nm}$
an increase of the contact area between the STM-tip and the semiconductor
sample results in a \textit{red shift} of the transition energy. The
explanation of this phenomenon is as follows. When the radius of the STM-tip
$R_{\mathrm{tip}}$ is much smaller than the thickness of the GaAs layer, the
electrostatic field from the STM-tip hardly extends into this layer, as shown
schematically in Fig. 5(a). Hence, the band bending in the GaAs layer is
negligible. In this case the energies of the transitions between hole and
electron levels are practically equal to those in the case without any
external field. An enlargement of the contact area increases the extension of
the electrostatic field into the structure, as shown in Fig. 5(b). This effect
is similar to that of an increase of the applied voltage. In both cases
transition energies become smaller (red shift). However, under the further
enlargement of the contact area the electrostatic field within this area
extends almost homogeneously over the whole semiconductor layer, see Fig.
5(c). As the luminescence originates in this case from the edge of the
STM-tip-induced quantum dot, there occurs a smaller rate of decrease of the
transition energies with increasing radius. Finally, when the radius
$R_{\mathrm{tip}}>12\ \mathrm{nm}$, the effects giving rise to the blue shift,
described in the previous paragraph, prevail.

Thus, we have shown a strong effect of the confinement potential, as
determined by the STM-tip radius and the applied voltage, on the
electroluminescence peak positions. Namely, while an increase of the voltage
on the STM-tip results in a \textit{red shift} of the electroluminescence
peaks \cite{my}, an increase of the contact area between the STM-tip and the
semiconductor structure results in a \textit{non-monotonous }variation of peak
positions. Below a certain critical value of the tip radius, the transition
energies decrease with increasing radius, while above the critical radius of
the tip, the transition energies are an increasing function of radius. These
results demonstrate vast possibilities of tuning optical characteristics of
the STM-tip induced quantum dot by varying parameters of the STM-tip.

\begin{acknowledgments}
This work has been supported by GOA BOF UA 2000, IUAP, FWO-V projects Nos.
G.0274.01, G.0435.03, WOG WO.025.99 (Belgium) and the European Commission
GROWTH Programme, NANOMAT project, contract No. G5RD-CT-2001-00545. The
research of M.K. has been made possible by a fellowship of the Royal
Netherlands Academy of Arts and Sciences.
\end{acknowledgments}

\newpage

\centerline{ \textbf{FIGURE CAPTIONS}}

\vspace{1cm}

\begin{enumerate}
\item[Fig. 1:] Scheme of the tip/insulator/GaAs/Al$_{0.25}$Ga$_{0.75}%
$As/In$_{0.25}$Ga$_{0.75}$As quantum well structure. The shown hole density
distribution is calculated for the STM-tip voltage $-3.1\ \mathrm{V}$ and the
radius of the tip $20~\mathrm{nm}$.

\item[Fig. 2:] Flow diagram of the numerical calculations.

\item[Fig. 3:] Positions of peaks in the electroluminescence spectra,
calculated for the STM-tip-induced quantum dot at $4.2~\mathrm{K}$,
$R_{\mathrm{tip}}=12~\mathrm{nm}$, versus bias $V_{\mathrm{tip}}$. Inset:
Typical electroluminescence spectrum calculated for the voltage on the STM-tip
$V_{\mathrm{tip}}=-3.1~\mathrm{V}$ and the STM-tip radius $R_{\mathrm{tip}%
}=12~\mathrm{nm}$.

\item[Fig. 4:] Positions of the electroluminescence peaks from Fig. 3 as a
function of the STM-tip radius at the voltage on the STM-tip $V_{\mathrm{tip}%
}=-3.1~\mathrm{V}$.

\item[Fig. 5:] Sketch of the electrostatic potential profile for the STM-tip
induced quantum dots with small (a), intermediate (b) and large (c) radius.
Darker regions correspond to a larger absolute value of the electrostatic potential.
\end{enumerate}

\newpage

\begin{figure}[ptb]
\centering   \includegraphics[scale=1.0]{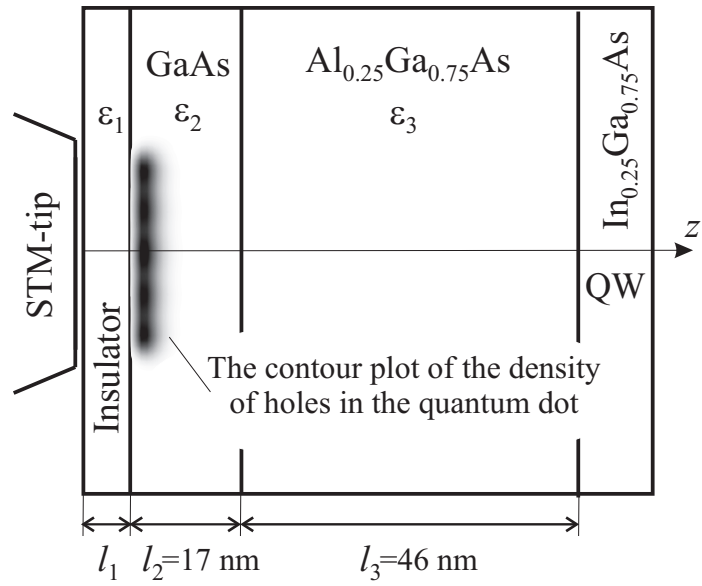}\caption{M.~D.~Croitoru,
V.~N.~Gladilin, V.~M.~Fomin, J.~T.~Devreese, M.~Kemerink, P.~M.~Koenraad,
J.~H.~Wolter}%
\label{fig1}%
\end{figure}

\begin{figure}[ptb]
\centering   \includegraphics[scale=1.0]{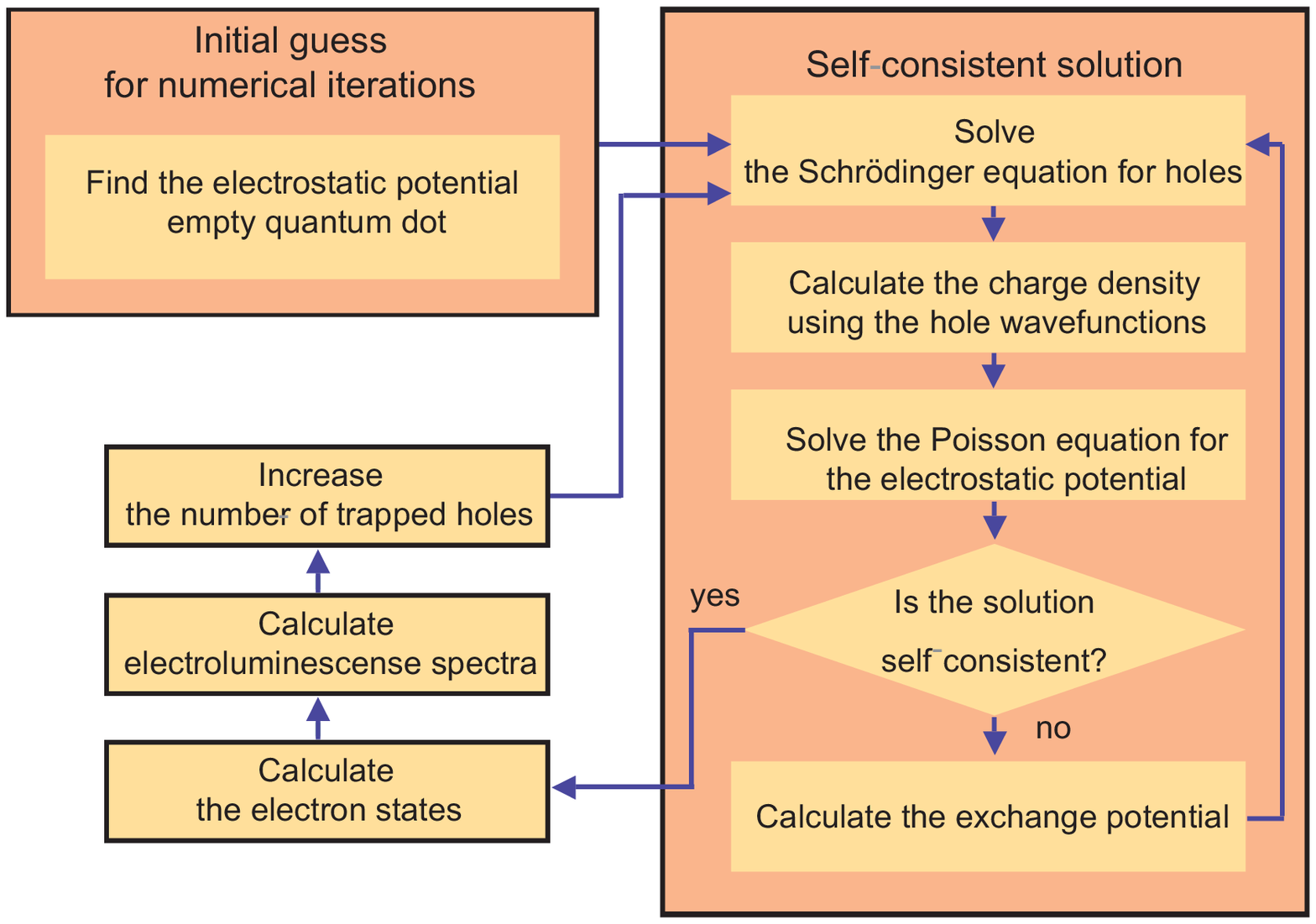}\caption{M.~D.~Croitoru,
V.~N.~Gladilin, V.~M.~Fomin, J.~T.~Devreese, M.~Kemerink, P.~M.~Koenraad,
J.~H.~Wolter}%
\label{fig2}%
\end{figure}

\begin{figure}[ptb]
\centering   \includegraphics[scale=1.4]{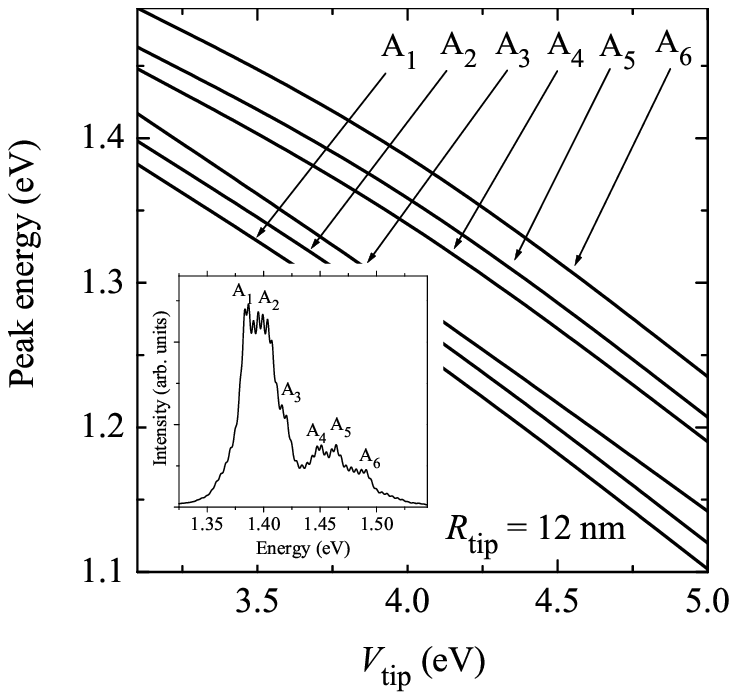}\caption{M.~D.~Croitoru,
V.~N.~Gladilin, V.~M.~Fomin, J.~T.~Devreese, M.~Kemerink, P.~M.~Koenraad,
J.~H.~Wolter}%
\label{fig3}%
\end{figure}

\begin{figure}[ptb]
\centering \includegraphics[scale=1.4]{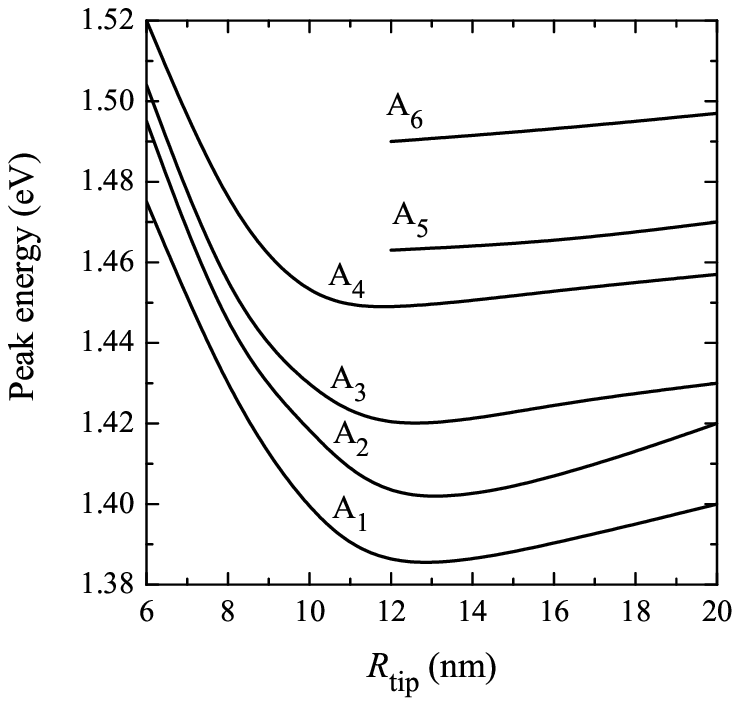}\caption{M.~D.~Croitoru,
V.~N.~Gladilin, V.~M.~Fomin, J.~T.~Devreese, M.~Kemerink, P.~M.~Koenraad,
J.~H.~Wolter}%
\label{fig4}%
\end{figure}

\begin{figure}[ptb]
\centering\includegraphics[scale=0.9]{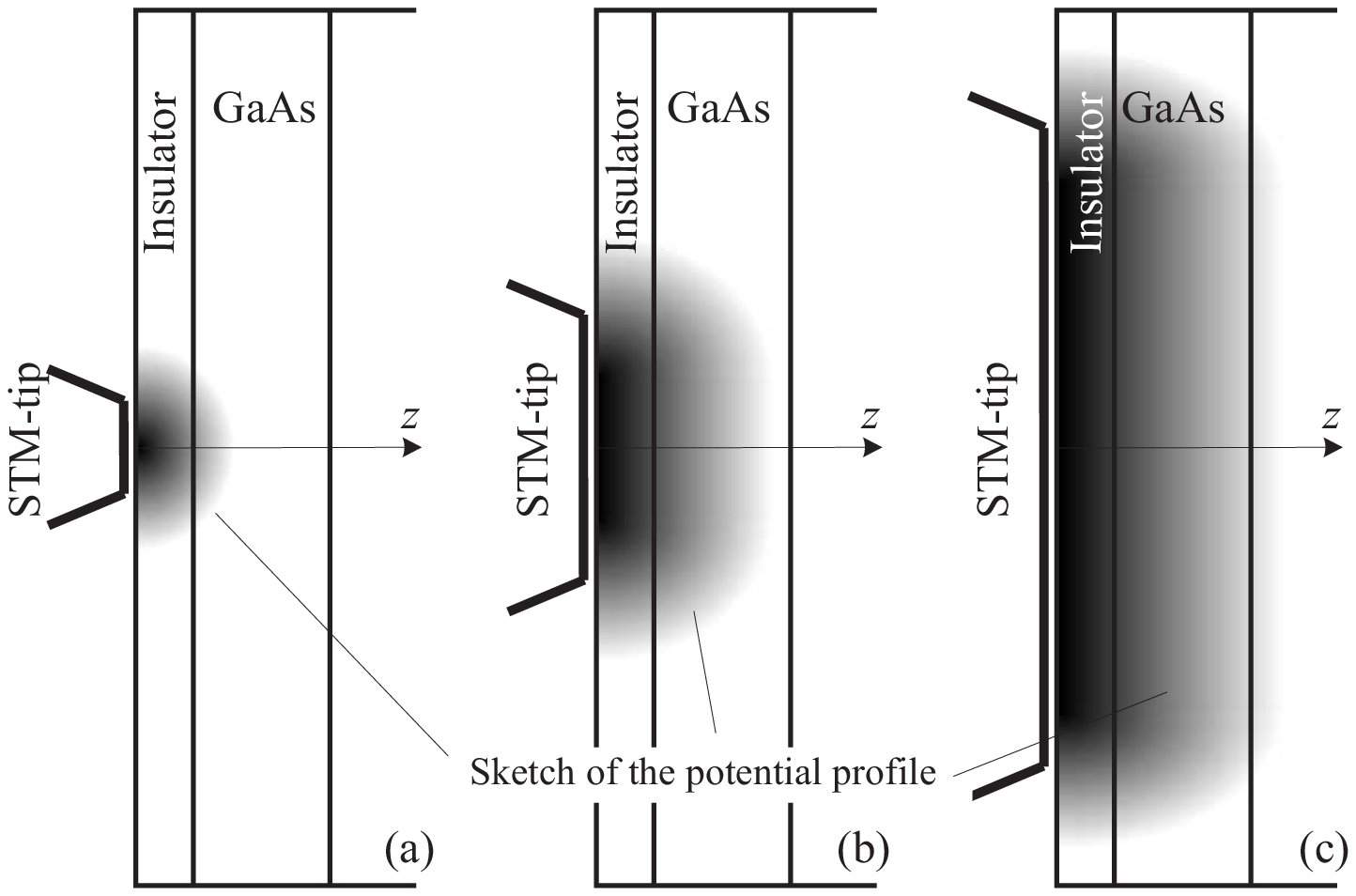}\caption{M.~D.~Croitoru,
V.~N.~Gladilin, V.~M.~Fomin, J.~T.~Devreese, M.~Kemerink, P.~M.~Koenraad,
J.~H.~Wolter}%
\label{fig5}%
\end{figure}

\end{document}